\documentclass[runningheads]{llncs}
\usepackage{amssymb}
\usepackage{listings}
\usepackage[usenames,dvipsnames]{xcolor}
\usepackage{textcomp}  
\usepackage{languages}
\usepackage{enumitem}
\usepackage{graphics}
\usepackage{ifthen}
\usepackage{natbib}
\usepackage[T1]{fontenc}  
\newcommand{\pair}[2]%
{\ensuremath{\langle}\code{#1}, \code{#2}\ensuremath{\rangle}}
\newboolean{showcomments}
\setboolean{showcomments}{true}
\ifthenelse{\boolean{showcomments}}
  {\newcommand{\nb}[2]{
    \fbox{\bfseries\sffamily\scriptsize#1}
    {\sf\small$\blacktriangleright$\textit{#2}$\blacktriangleleft$}
   }
  }
  {\newcommand{\nb}[2]{}
  }

\begin{document}
\pagestyle{headings}  
\mainmatter
\title{The Essence of Inheritance}
\titlerunning{Essence of Inheritance}  
%
\author{Andrew~P.~Black\inst{1} \and Kim~B.~Bruce\inst{2}
\and R.~James~Noble\inst{3}}
\authorrunning{Black, Bruce, and Noble} 
%
\institute{Portland State University, Oregon, USA,\\
\email{black@cs.pdx.edu},
\and
Pomona College, Claremont, California, USA,\\
\email{kim@cs.pomona.edu}
\and
Victoria University of Wellington, New Zealand
\email{kjx@ecs.vuw.ac.nz}
}
\maketitle              
\begin{abstract}
Programming languages serve a dual purpose: to communicate programs to computers, and to communicate programs to humans.
Indeed, it is this dual purpose that makes programming language design 
a constrained and challenging problem.
Inheritance is an \textit{essential} aspect of 
that second purpose: it is a tool to improve communication.
Humans understand new concepts most readily
by \emph{first} looking at a number of concrete examples, 
and \emph{later} abstracting over those examples.
The essence of inheritance is that it mirrors this process: it provides a formal mechanism for
moving from the concrete to the abstract.
\end{abstract}

\section{Introduction}

Shall I be abstract or concrete?

An abstract program is more general, and thus has greater potential to be
reused.
However, a concrete program will usually solve the specific problem at hand 
more simply.

One factor that should influence my choice is the ease with which a program can be understood.  Concrete programs ease understanding by making manifest the action of
their subcomponents.   But, sometimes a seemingly small change may require a concrete
program to be extensively restructured, when judicious use of abstraction would have 
allowed the same change to be made simply by providing a different argument.

Or, I could use inheritance.

The essence of inheritance is that it lets us avoid the unsatisfying choice between abstract and concrete.
Inheritance lets us start by writing a concrete program, and then later on abstracting over 
a concrete element.
This abstraction step is \emph{not} performed by editing the concrete program to introduce a new parameter.
That is what would be necessary without inheritance.
To the contrary: inheritance allows us to treat the concrete element 
\emph{as if it were a parameter}, without actually changing the code.
We call this ex post facto parameterization; we will illustrate the process with examples in Sections~\ref{interp} and \ref{OTP}. 

Inheritance has been shunned by the designers of functional languages.  
Certainly, it is a difficult feature to specify precisely, and to implement efficiently, 
because it means (at least in the most general formulations) 
that any apparent constant might, once inherited, become a variable.
But, as Einstein is reputed to have said, in the middle of difficulty lies opportunity.
The especial value of inheritance is as an aid to program 
understanding.
It is particularly valuable where the best way to understand a
complex program is to start with a simpler one and approach
the complex goal in small steps.


Our emphasis on the value of inheritance as an aid to human understanding,
rather than on its formal properties, is deliberate, and long overdue.
Since the pioneering work of~\citet*{cook1989a}, it has been clear that,
from a formal point of view, inheritance is equivalent to parameterization.
This has, we believe, caused designers of functional languages to regard 
inheritance as unimportant, unnecessary, or even undesirable, arguing (correctly)
that it can be simulated using higher-order parameterization.
This line of argument misses the point that two formally-equivalent mechanisms may behave quite differently with respect to human cognition.

It has also long been known that \textit{Inheritance is Not Subtyping}~\citep{cook1990}.
In spite of this, many programming languages conflate subtyping and inheritance; 
Java, for example, restricts the use of inheritance so that the inheritance
hierarchy is a sub-tree of the type hierarchy.
Our goal in this paper is to consider inheritance as a mechanism in its own right, 
quite separate from the subtyping relation.
We are aided in this goal by casting our example in the Grace programming language~\citep{black2012}, 
which cleanly separates inheritance and subtyping.
 
The form and content of this paper are a homage to Wadler's ``Essence of Functional Programming''~\citeyearpar{wadler1992a}, which was itself inspired by Reynold's ``Essence of ALGOL''~\citeyearpar{reynol1981}.
The first example that we use to illustrate the value of inheritance, discussed in Section~\ref{interp}, 
is based on Wadler's interpreter for a simple language, which is successively extended to include additional features.
(\citet{reynol1972} had previously defined a series of interpreters in a classic paper that 
motivated the use of continuations.)
The second example is based on Armstrong's description of the Erlang OTP Platform~\citep[Ch~16]{armstr2007}; this is discussed in Section~\ref{OTP}.
We conclude in Section~\ref{discussion}, where we also discuss some
of the consequences for type-checking.

\section{Interpreting Inheritance}
\label{interp}
\choosegrace

This section demonstrates the thesis that inheritance enhances understandability, 
by presenting several variations of an implementation of 
a fragment of an expression language.
\citet{wadler1992a} uses an interpreter for the lambda calculus, 
but we use a simpler language inspired by the ``First Monad Tutorial''~\citep{wadler2013}.

We show our examples in Grace~\citep{black2012}, a simple object-oriented language designed
for teaching, but no prior knowledge of the language is assumed.
What the reader will require is a passing familiarity with the basics of object-oriented programming; 
for general background, see~\cite{cox1986}, or \cite{Blac09a}.
Like Haskell, Grace does not require the programmer to specify 
types on every declaration, although, also like Haskell, 
the programmer may well find that adding type annotations makes it
easier to find errors.

\subsection{Variation Minus One: Object-Oriented Evaluation} 
\label{simple}

We start with a simple object-oriented evaluator
for Wadler's tutorial example~\citeyearpar{wadler2013}\,---\,mathematical expressions consisting
only of constants \lstinline+con+ and division \lstinline+div+\,---\,so
\lstinline+1/2+ is represented as
\lstinline+div(con 1, con 2)+.
The object-oriented style integrates the data and the descriptions of the computations on that data.
Thus, there is no separation between the ``interpreter''
and the ``terms'' that it interprets;
instead the object-oriented program contains ``nodes'' of various kind, 
which represent \emph{both} expressions \emph{and} the rules for their evaluation.

In Grace,
\lstinline+con+ and \lstinline+div+ are classes; they correspond to
Wadler's algebraic data constructors.
Grace classes (like those in O'Caml and Python) are
essentially methods that return new objects.
Evaluation of an expression in the example language is accomplished by requesting
that the objects created by these classes execute their own
\lstinline+eval+ methods, rather than by invoking a globally-defined function \code{interp}.
The resulting representation of \code{simpleExpressions} is straightforward.
In addition to the \code{eval} methods, the objects are also given \code{asString}
methods that play a role similar to that of Wadler's \code{show}$*$ functions:
they return strings for human consumption.
(Braces in a string constructor interpolate values into the string.)

\inputgrace{simpleExpressions}{minigrace}

Given these definitions, we can instantiate an expression tree,
evaluate it, and print out the result. (Grace's \lstinline+def+ is
similar to ML and Scala's \textsf{\bfseries let}). 

\begin{lstlisting}
import "simpleExpressions" as s
def e = s.simpleExpressions
def answer = e.div(e.div(e.con 1932, e.con 2), e.con 23)
print (answer.eval)         // prints 42
def error = e.div(e.con 1, e.con 0)
print (error.eval)         // prints infinity
\end{lstlisting} 
\noindent

\citet{wadler1992a} does not present code equivalent to this variation, which is why we have labelled it ``Variation Minus One''. Instead, he starts with ``Variation Zero'', which is a monadic interpreter based on a trivial monad.

\subsection{What is a Monad?}
For our purposes, a monad is an object that encapsulates an effectful computation.
If the underlying computation produces an answer \code{a}, then the 
monad can be thought of as ``wrapping'' the computation of \code{a}.
If you already understand monads, you should feel free to skip to Section~\ref{monadic}.

All monad objects provide a ``bind'' method, which, following Haskell, we spell \code{>>=}.
(In Grace, binary methods can be named by arbitrary sequences of
operator symbols; there is no precedence.)
In general, a monad will also encapsulate some machinery specialized to its particular purpose
computations that execute within it.
For example, the output monad of Section~\ref{outputmonad} encapsulates a string
and an operation by which a computation can append its output.

The bind method \code{>>=} represents sequencing.
The right-hand argument of \code{>>=} is a Grace code block, 
which denotes a $\lambda$-expression or function, and which represents the
computation that should succeed the one in the monad.
(The successor computation is like a continuation; \citet[\S3]{wadler1992a} explores the similarities and differences.)
The request \code!m>>= { a -> ... }! asks the monad object \code{m} to build a new
monad that, when executed, will first
execute \code{m}'s computation, 
bind the answer to the parameter \code{a} of the block, 
and then execute the successor computation represented by \code{...}.
Think of the symbol \code{>>=} as piping the result of the computation to its left into
the function object to its right.
The block should answer a monad, so the type of \code{>>=} is 
\mbox{\code{(Block1<A, Monad>) -> Monad}}. (The Grace type \mbox{\code{Block1<X, Y>}} 
describes a $\lambda$-expression that takes an argument of 
type \code{X} and returns a result of type \code{Y}.)
The monad type is actually parameterized by the type of its contents, but
Grace allows us to omit type parameters, which we do here for simplicity.
(In Grace's gradual type system, 
omitted types and type parameters are assumed to be \code{Unknown},
which means that we are choosing to say nothing about the type of the corresponding object.)

In the object-oriented formulation of monads, the monad classes construct monad objects, and thus subsume the operation that Haskell calls  
\choosehaskell
\code{unitM}\,or\,\mbox{\code{return}.}
\choosegrace
In the sections that follow, we will introduce four monad classes:
\mbox{\code{pure(contents)}}, which encapsulates pure computations 
that have no effects;
\code{raise(reason)}, which encapsulates computations that raise exceptions;
\code{stateMonad(contents)}, which encapsulates stateful computations;
and \code{monad(contents)output(str)}, which encapsulates computations that
produce output.
The name of this last class is an example of a method name with multiple parts,
a Grace feature designed to improve readability that is similar to Smalltalk
method selectors.
The class \code{monad()output()} expects two argument lists, each with a single argument.

Informally, the monad classes get us into a monad, and \code{>>=} gets us around the monad.  
How do we get out of the monad? 
In general, such operations require a more \textit{ad hoc} design.
In this article, it will suffice to provide methods \code{asString} 
on the monads that answer strings.
\subsection{Variation Zero: Monadic Evaluation} 
\label{monadic}
We now \emph{adapt} the code of Section~\ref{simple} to be monadic.
Here we see the key difference between an approach that uses
inheritance and one that does not: rather than define the new
expression evaluation mechanism from scratch, we make a new \lstinline+monadicExpression+
module that uses inheritance to modify the components
imported from the previous \lstinline+simpleExpressions+ module.


For a monadic interpreter, we of course need a monad, which we 
define in Grace as a class called \code{pure}, since 
it encapsulates effect-free computations that answer
pure values.  
The bind method
\code{>>=} applies its parameter \code{k} to the contents of the monad; 
in Grace, application of a $\lambda$-expression is performed explicitly using the method \code{apply}.

\inputgrace{monadicExpressions}{minigrace}

Having established this framework, we then define
the nodes of this monadic variant of expressions;
these differ from the simple ones of \S\,\ref{simple} \emph{only}
in the \code{eval} method, so the \emph{only} content of these new classes 
are \code{eval} methods that override the inherited ones.
The \code{inherits} statement on line 15 says that the methods of \code{con}
in \code{monadicExpressions} are the same as those of \code{con} in \code{simpleExpressions},
\emph{except} for the overridden method \code{eval}.
The \code{alias} clause on the \code{inherits} statement provides 
a new name for the inherited \code{eval}, so that the overriding method body can 
lift the value produced by the inherited \code{eval} method
into the monad.
The \code{eval} method of \code{div} is similar.
We ask readers concerned about type-checking this instance of ``the expression problem'' 
to suspend disbelief until Section~\ref{types}.

Notice that the original definition of \code{se.simpeExpressions} in Section~\ref{simple}
did not specify that \code{eval} was a parameter.  
On the contrary: it was specified as a simple concrete method.
The method \code{eval} becomes a parameter only when an inheritor of \code{se.simpeExpressions} chooses to
override it.  
This is what we meant by \textit{ex post facto parameterization} in the introduction.


Here is a test program parallel to the previous one: the output indicates
that the results are in the monad.

\begin{lstlisting}
import "monadicExpressions" as m
def e = m.monadicExpressions
def answer = e.div(e.div(e.con 1932, e.con 2), e.con 23)
print (answer.eval)         // prints pure(42)
def error = e.div(e.con 1, e.con 0)
print (error.eval)         // prints pure(infinity)
\end{lstlisting}


In the remainder of this section, we show how three of Wadler's variations
on this monadic interpreter\,---\,for exceptions, state, and output\,---\,can be 
implemented incrementally using inheritance.
\subsection{Variation One: Evaluation with Exceptions}

The basic idea behind adding exceptions is that the result of an 
evaluation is no longer always a simple value (like 42)
but may also be an exception.  
Moreover, exceptions encountered during the
evaluation of an expression terminate the evaluation and emit the
exception itself as the answer.
In our toy language, the only source of exceptions will be 
division by zero.


We have to extend our monad to deal with exceptional results.
Working in Haskell, Wadler redefines
the algebraic data type contained in the monad, starting from scratch.
Working in Grace, we define
\code{exceptionExpressions} by inheriting from
\mbox{\code{monadicExpressions},} and need write code only for the differences.
We start by defining a new monad class \lstinline+raise+ (lines 5--8),
whose bind method stops execution when an exception is encountered
by immediately returning \lstinline+self+, thus discarding any 
computations passed to it as its parameter \code{k}. 

\inputgrace{exceptionExpressions}{minigrace}


We must also change the \code{eval} methods to raise exceptions at the appropriate time.
Since the evaluation of constants cannot raise an exception, the \code{con}
class is unchanged, \emph{so we say nothing}, and use the inherited \code{con}.
In contrast, evaluating a \code{div} can raise an exception, so we have
to provide a new \code{eval} method for \code{div}, shown on lines 11--16.
This code returns a value in the \code{raise} monad 
when the divisor is zero.

If you are familiar with Wadler's addition of error messages to his monadic
interpreter~\citep[\S2.3]{wadler1992a} this will look quite familiar.  
In fact, the developments are so similar that it is easy to overlook the differences:  
\begin{enumerate}[topsep=1.5ex]
	\item  At the end of his description of the changes necessary to introduce error messages,
	Wadler writes: 
		``To modify the interpreter, substitute monad \code{E} for monad \code{M}, 
		and replace each occurrence of \code{unitE Wrong} by a suitable call to 
		\code{errorE}. \ldots{} No other changes are required.''
	Wadler is giving us editing instructions!  In contrast, the box above represents a file of
	real Grace code that can be compiled and executed.  It contains, if you will, 
	not only the new definitions for the \code{eval} method and the monad, but 
	\emph{also the editing instructions} required to install them.

	\item  Not only does the boxed code represent a unit of compilation, it also represents a 
	\emph{unit of understanding}.   We believe that it is easier to understand a complex 
	structure like a monad with exceptions by first understanding the monad \code{pure}, 
	and \emph{then} understanding the exceptions.   
	Perhaps Wadler agrees with this, for he himself uses just this form of exposition in his \emph{paper}. 
	But, lacking inheritance, he cannot capture this stepwise exposition in his \emph{code}.
\end{enumerate}
\subsection{Variation Three: Propagating State} 
\label{monadicVariationThree}
To illustrate the manipulation of state, Wadler keeps a count of 
the number of reductions; we will count the number of divisions.
Each computation is given an input state in which to execute, 
and returns a potentially different output state; 
the difference between the input and output states reflect the changes to 
the state effected by the computation. 
Rather than representing $\langle \mathit{result}, \mathit{state}\rangle$ pairs with 
anonymous tuples, we will use \code{Response} objects with two methods, shown below.
\inputgrace{response}{minigrace}
What changes must be made to \code{monadicExpressions} to support state?
The key difference between the \code{stateMonad} and the \code{pure} monad is in the \code{>>=} method. We follow the conventional
functional approach, with the contents of the monad being a function that, when applied to a state, returns
a \code{Response}.
The method \code{executeIn} captures this idea. 

\inputgrace{statefulEvaluation}{minigrace}

The \code{>>=} method of the state monad (lines 9--13) returns a new state monad that,
when executed, will first execute its contents in the given state \code{s}, 
then apply its argument \code{k}
to the \code{result}, and then execute the contents of that monad in the threaded state.
Given this definition of \code{>>=}, and a redefinition of the
\lstinline+pure+ method to ensure that values are lifted into the correct
monad, we do not need to redefine \emph{any} of the tree nodes\,---\,so long
as their evaluation does not involve using the state.

Of course, the \emph{point} of this extension is to enable some evaluations to depend on state.
We will follow Wadler in using state in a somewhat artificial way: 
the state will be a single number that counts the number of divisions.
The \lstinline+tally+ method increments the count stored in the state 
and answers \lstinline+done+ (Grace's
unit), indicating that its purpose is to cause an effect rather than to compute a result.


Naturally, counting divisions requires extending the \code{eval} operation in the \code{div}
class (but not elsewhere)
to do the counting.
The overriding \code{eval} method on lines 24--26 first counts using \code{tally}, 
and then calls the inherited
\code{eval} method.  Notice also that the \mbox{\code{asString}}
operation of the monad (line 7) executes
the contents of the monad in the state \code{0}, representing 
the initial value of the counter.
This is exactly parallel to Wadler's function \code{showS}.

\pagebreak
Here is an example program similar to the previous one:
\begin{lstlisting}
import "statefulEvaluation" as se
def e = se.statefulEvaluation

def simple = e.con 34
print (simple.eval)         // prints stateMonad(result(34) state(0))

def answer = e.div(e.div(e.con 1932, e.con 2), e.con 23)
print (answer.eval)         // prints stateMonad(result(42) state(2))

def error = e.div(e.con 1, e.con 0)
print (error.eval)         // prints stateMonad(result(infinity) state(1))
\end{lstlisting}
\subsection{Variation Four: Output} 
\label{outputmonad}

Our final example builds up an output string, using similar techniques.
We reuse \code{monadicExpressions}, this time
overriding the inherited \code{eval} methods to produce output.
We provide a new monad that holds both result and
output, and is equipped with a bind operator that
accumulates the output at each step.

\inputgrace{outputEvaluation}{minigrace}
\pagebreak
\noindent
Here is an example program, with its output.
\begin{lstlisting}
import "outputEvaluation" as o

def e = o.outputEvaluation
def answer = e.div(e.div(e.con 1932, e.con 2), e.con 23)
print (answer.eval)
// prints output monad contents: 42 output:
// con 1932 = 1932
// con 2 = 2
// div(con 1932, con 2) = 966
// con 23 = 23
// div(div(con 1932, con 2), con 23) = 42
def error = e.div(e.con 1, e.con 0)
print (error.eval)

// prints output monad contents: infinity output:
// con 1 = 1
// con 0 = 0
// div(con 1, con 0) = infinity
\end{lstlisting}

\section{The Erlang OTP Platform}
\label{OTP}

The essential role that inheritance can play in \emph{explaining} how a software system
works was brought home to Black in 2011. 
Black was on sabbatical in Edinburgh, hosted by Phil Wadler.
Black and Wadler had been discussing ways of characterizing encapsulated state\,---\,principally monads and effect systems.
Erlang came up in the conversation as an example of a language in which state is handled 
explicitly, and Black started studying Armstrong's \textit{Programming Erlang}~\citeyearpar{armstr2007}.

The Erlang OTP ``generic server'' provides properties such as transactions, 
scalability, and dynamic code update for arbitrary server behaviours.
Transactions can be implemented particularly simply because state is explicit.
Armstrong writes:
``Put simply, the [generic server] solves the nonfunctional parts of the 
problem, while the callback solves the functional part.  
The nice part
about this is that the nonfunctional parts of the problem 
(for example, how to do live code upgrades) 
are the same for all applications.''~\citep[p.\,286]{armstr2007}

A reader of this section of Armstrong's book can't help but get the sense
that this material is significant. 
In a separate italicized paragraph, Armstrong writes:
``\hspace{0.3pt}This is the most important section in the entire book, 
so read it once, read it twice, read it 100 times\,---\,just make sure the message sinks in.''

The best way that Armstrong can find to explain this most-important-of-messages is to write a small server program in Erlang, 
and then generalize this program to add first transactions,
and then hot code swapping.
The best way that Black could find to understand this message was to 
re-implement Armstrong's server in another language\,---\,Smalltalk.
Using Smalltalk's inheritance, he was able to reflect Armstrong's development,
not as a series of \emph{separate} programs, but in the stepwise development of a \emph{single} program.

\subsection{Armstrong's Goal}

\lstset{language=grace} 
What does Armstrong seek to achieve with the generic server?
In this context, a ``server'' is a computational engine with access to persistent memory.
Servers typically run on a remote computer, and in the Erlang world the primary server is backed-up by a secondary server that takes over if the primary should fail.
For uniformity with the proceeding example, we here present a simplified version
of the OTP server in Grace;  in addition to Armstrong's Erlang version, 
our code is also based on  \citeauthor*{bierma2008}'s  
translation of the OTP server to a Java-like language~\citeyearpar{bierma2008}.
For brevity, and to focus attention on the use of inheritance, our version omits name resolution, remote requests, concurrency and failover.
 
To be concrete, let's imagine two particular servers: a ``name server'' 
and a ``calculation server''.
The name server remembers the locations of objects, with interface:
\begin{grace}
type NameServer = {
    add(name:String) place(p:Location) -> Done
    whereIs(name:String) -> Location
}
\end{grace}
\noindent
The name of the first method in the above type is \code{add()place()},
a two-part name with two parameters.

The calculation server acts like a one-function calculator with a memory;
\code{clear} clears the memory, and \code{add} 
adds its argument to the memory, and stores the result in the memory as well
as returning it.
The calculation server has the interface 
\begin{grace}
type CalculationServer = {
    clear -> Number
    add(e:Number) -> Number
}
\end{grace}
Both of these servers maintain state; we will see later why this is relevant.

Armstrong refers the actual server code as ``the callback''.  
His goal is to write these callbacks in a simple sequential style, but with explicit state.
The generic
server can then add properties such as transactions, failover and hot-swapping. 
The simple sequential implementation of the name server is shown on the following page.

The state of this ``callback'' is represented by a \code{Dictionary} object that stores the 
\textit{name} $\rightarrow$ \textit{location} mapping.  
The method \code{initialState} returns the initial state: a new, empty, \code{Dictionary}.
The method \code{add()place()state()} is used to implement the client's \code{add()place()}
method. 
The generic server provides the additional state argument, an object representing the callback's state.
The method returns a \code{Response} (as in Section~\ref{monadicVariationThree})
comprising the \code{newState} dictionary, and the actual result of the operation, \code{p}.
Similarly, the method \code{whereIs()state()} is used to implement the client's \code{whereIs()} method.
The generic server again provides the additional state argument.  This method
returns a \code{Response} comprising the result of looking-up \code{name} in
\code{dict} and the (unchanged) state.

\inputgrace{nameServer}{minigrace}

Finally, let's consider what happens if this name server callback is asked for the location of a name that is not in the dictionary.
The lookup \code{dict.at(name)} will raise an exception,
which the callback itself does not handle.

Notice that our \textit{nameServer} module contains a class \code{callback} 
whose instances match the type
\begin{grace}
type Callback<S> = type {
	initialState -> S
}
\end{grace}
for appropriate values of \code{S}. This is true of all server callback modules.
Particular server callbacks extend this type with additional methods, all of which have a name that
ends with the word \code{state}, and which take an extra argument of type \code{S} that represents their state.

\subsection{The Basic Server}
Our class \code{server} corresponds to Armstrong's module \code{server1}.
This is the ``generic server'' into which is installed the ``callback'' that programs it to provide a particular function (like name lookup, or calculation).

A \code{Request} encapsulates the name of an operation 
and an argument list.
The basic server implements two methods: \code{handle()}, which processes a single 
incoming request, and \code{serverLoop()}, which manages the request queue.
\inputgrace{basicServer}{minigrace}

A \code{server} implements three methods.  Method \code{startUp(name)} loads the
callback module \code{name} and initializes the server's state to that required by the
newly-loaded callback.

The method \code{handle} accepts an incoming request, such as ``add()place()'', 
and appends the string ``state'', to obtain method name like  ``add()place()state''.
It then requests that the callback executes its method with this name, passing it the
arguments from the request and an additional state argument.
Thus, a request like 
\code{add "BuckinghamPalace" place "London"}
might be transmitted to the nameServer as 
\mbox{\code{add "BuckinghamPalace"}} \mbox{\code{ place "London"}} 
\mbox{\code{state (dictionary.empty)}}.
The state component of the response would then be a dictionary containing the mapping from
\code{"BuckinghamPalace"} to \code{"London"}; this new dictionary would
provide the state argument for the next request.

The method \code{serverLoop} is a simplified version of Armstrong's \code{loop/3} that omits
the code necessary to receive messages and send back replies, and instead uses a local queue of messages and Grace's normal method-return mechanism.

Here is some code that exercises the basic server:

\begin{lstlisting}
import "basicServer" as basic

class request(methodName)withArgs(args) {
    method name { methodName }
    method arguments { args }
}
def queue = [
    request "add()place()" withArgs ["BuckinghamPalace", "London"],
    request "add()place()" withArgs ["EiffelTower", "Paris"],
    request "whereIs()" withArgs ["EiffelTower"]
]
print "starting basicServer"
basic.server("nameServer").serverLoop(queue)
print "done"
\end{lstlisting}

\sloppypar
To keep this illustration as simple as possible, this code constructs the requests explicitly; 
in a real remote server system, the requests would be constructed using reflection, or an 
RPC stub generator.  This is why they appear as, for example, \mbox{\code{request "add()place()" withArgs ["BuckinghamPalace", "London"]}}, 
instead of as \mbox{\code{add "BuckinghamPalace"}} \mbox{\code{ place "London"}}.
Here is the log output:
\begin{quote}
\it\small
starting basicServer\\
handle: add()place() args: [BuckinghamPalace, London]\\
\hspace*{2em}
    result: London\\
handle: add()place() args: [EiffelTower, Paris]\\
\hspace*{2em}
    result: Paris\\
handle: whereIs() args: [EiffelTower]\\
\hspace*{2em}
    result: Paris\\
done
\end{quote}
Note that if the server callback raises an exception, it will crash the whole server.

\subsection{Adding Transactions}
Armstrong's \code{server2} adds transaction semantics: if the
requested operation raises an exception, the server loop continues with
the \emph{original} value of the state.  
In contrast, if the requested operation completes normally, the server continues with the
\emph{new} value of the state.

Lacking inheritance, the only way that Armstrong can explain this to his
readers is to present the \emph{entire text} of a new module, \code{server2}.
The reader is left to compare each function in \code{server2} 
with the prior version in \code{server1}.
The \code{start} functions seem to be identical; 
the \code{rpc} functions are similar, 
except that the \code{receive} clause in \code{server2} has been extended to accommodate
an additional component in the reply messages.
The function \code{loop} seems to be completely different.

In the Grace version, the differences are much easier to find.
In the \textit{transactionServer} module, the class \code{server} is derived from
\textit{basicServer}'s \code{server} using inheritance:
\inputgrace{transactionServer}{minigrace}
The \code{handle} method is overridden, but
\emph{nothing else changes}. It's easy to see that the extent of the change
is the addition of the \code{try()catch()} clause to the \code{handle} method.

If we now try and make bogus requests:
\begin{lstlisting}
import "transactionServer" as transaction

class request(methodName)withArgs(args) {
    method name { methodName }
    method arguments { args }
}
def queue = [
    request "add()place()" withArgs ["BuckinghamPalace", "London"],
    request "add()place()" withArgs ["EiffelTower", "Paris"],
    request "whereIs()" withArgs ["EiffelTower"],
    request "boojum()" withArgs ["EiffelTower"],
    request "whereIs()" withArgs ["BuckinghamPalace"]
]
print "starting transactionServer"
transaction.server("nameServer").serverLoop(queue)
print "done"
\end{lstlisting}
they will be safely ignored:
\begin{quote}
\it\small
starting transactionServer\\
handle: add()place() args: [BuckinghamPalace, London]\\
\hspace*{2em}
    result: London\\
handle: add()place() args: [EiffelTower, Paris]\\
\hspace*{2em}
    result: Paris\\
handle: whereIs() args: [EiffelTower]\\
\hspace*{2em}
    result: Paris\\
Error --- server crashed with NoSuchMethod: no method boojum()state in mirror for a callback\\
handle: boojum() args: [EiffelTower]\\
\hspace*{2em}
    result: !CRASH!\\
handle: whereIs() args: [BuckinghamPalace]\\
\hspace*{2em}
    result: London\\
done
\end{quote}

Armstrong emphasizes that the same server callback can be run under both the basic server and the transaction sever.  
This is also true for the Grace version, but that's not what we wish to emphasize.
Our point is that inheritance makes it much easier to understand the critical 
differences between \textit{basicServer} and \textit{transactionServer} than does
rewriting the whole server, as Armstrong is forced to do.
\subsection{The Hot-Swap Server}

Armstrong's \code{server3} adds ``hot swapping'' to his \code{server1}; once again
he is forced to rewrite the whole server from scratch, and the reader must compare
the two versions of the code, line by line, to find the differences. 
Our Grace version instead adds hot swapping to the \code{transactionServer}, again using inheritance.
\inputgrace{hotSwapServer}{minigrace}
In \textit{hotSwapServer}, class \code{server} overrides the \code{handle} method
with a version that checks for the special request \code{!HOTSWAP!}.
Other requests are delegated to the \mbox{\code{handle}} method inherited from
\textit{transactionServer}.  
Once again, it is clear that \emph{nothing else changes}.

Now we can try to change the name server into a calculation server:
\begin{lstlisting}
import "hotSwapServer" as hotSwap

class request(methodName) withArgs(args) {
    method name { methodName }
    method arguments { args }
}
def queue = [
    request "add()place()" withArgs ["EiffelTower", "Paris"],
    request "whereIs()" withArgs ["EiffelTower"],
    request "!HOTSWAP!" withArgs ["calculator"],
    request "whereIs()" withArgs ["EiffelTower"],
    request "add()" withArgs [3],
    request "add()" withArgs [4]
]
print "starting hotSwapServer"
hotSwap.server("nameServer").serverLoop(queue)
print "done"
\end{lstlisting}
Here is the output:
\begin{quote}
\it\small
starting hotSwapServer\\
handle: add()place() args: [EiffelTower, Paris]\\
\hspace*{2em}
    result: Paris\\
handle: whereIs() args: [EiffelTower]\\
\hspace*{2em}
    result: Paris\\
handle: !HOTSWAP! args: [calculator]\\
\hspace*{2em}
    result: calculator started.\\
Error --- server crashed with NoSuchMethod: no method whereIs()state in mirror for a callback\\
handle: whereIs() args: [EiffelTower]\\
\hspace*{2em}
    result: !CRASH!\\
handle: add() args: [3]\\
\hspace*{2em}
    result: 3\\
handle: add() args: [4]\\
\hspace*{2em}
    result: 7\\
done
\end{quote}
\subsection{Summary}
In summary, we can say that what Armstrong found necessary to present as a 
series of completely separate programs, can be conveniently expressed in Grace
as one program that uses inheritance.
As a consequence, the final goal\,---\,a hot-swappable server that supports
transactions\,---\,cannot be found in a single, monolithic piece of code,
but instead is a composition of three modules.
But far from being a problem, this is an \emph{advantage}.  
The presentation using inheritance lets each feature be implemented, 
and understood, separately.   
Indeed, the structure of the code mirrors quite closely the structure of 
Armstrong's own exposition in his book.

\section{Discussion and Conclusion}
\label{discussion}
\lstset{language=grace}
We will be the first to admit that a few small examples prove nothing.
Moreover, we are aware that there are places where our argument needs improvement.  Here we comment on two of these.

\subsection{From the Abstract to the Concrete}
\label{Iterable}
Inheritance can be used in many ways other than the way illustrated here.
Our own experience with inheritance is that we sometimes start out with a concrete class $X$,
and then write a new concrete class $Y$ doing something similar.
We then abstract by creating an (abstract) superclass $A$ that contains the attributes
that $A$ and $B$ have in common, and rewrite $X$ and $Y$ to inherit from $A$.

When programmers have a lot of experience with the domain, 
they might even \emph{start out} with the abstract superclass.
We did exactly this when building the Grace collections library, which contains
abstract superclasses like \code{iterable}, 
which can be inherited by any concrete class that defines an \code{iterator} method.
Syntax aside, an abstract superclass is essentially a
functor returning a set of methods, and the abstract method(s) of the superclass
are its \emph{explicit} parameters.  
Used in this way, inheritance is little more than a convenient parameterization construct.

If we admit that this use of inheritance as a parameterization construct is common, 
the reader may wonder why
have we used the bulk of this paper to emphasize the  
use of inheritance as a construct for differential programming.
The answer is that we wish to capture the \emph{essence} of inheritance.
According to \textit{The Blackwell Dictionary of Western Philosophy},
``essence is the property of a thing without which it could not be what it is.''~\citep[p.223]{bunnin2004}.
In this sense, the essence of inheritance is its ability to override 
a concrete entity, and thus effectively turn a constant into a parameter.
There is nothing at all wrong with using inheritance to share abstract entities designed
for reuse, that is, to move from the abstract to the concrete.
But what makes inheritance unique is its ability to create parameters out of 
constants\,---\,what we have called ex post facto parameterization.

\subsection{What About Types?}
\label{types}

Anyone who has read both Wadler's ``The Essence of functional programming'' 
and our Section~\ref{interp} can hardly fail to notice that Wadler's code is rich with 
type information, while our code omits many type annotations.  This prompts the question:
can the inheritance techniques that we advocate be well-typed?

The way that inheritance is used in Section~\ref{interp}\,---\,overriding the 
definition of the monad in descendants\,---\,poses some fundamental difficulties for 
typechecking.  
As the components of the expression class are changed to add new functionality,
the types that describe those components also change.  
For example, the parameters \code{l} and \code{r} of the class \code{div} in module ``simpleExpressions'' have \code{eval} methods that answer numbers, 
whereas the corresponding parameters to \code{div} in module 
``monadicExpressions'' have \code{eval} methods that answer monads.
If there were other methods that used the results of \code{eval}, 
these would also need to be changed to maintain type correctness.
Even changing a type to a subtype is not safe in general, because the type
may be used to specify a requirement on a parameter as well as a property of a result.

Allowing an overriding method in a subclass to have a type different from
that of the overridden method clashes with one of the
sacred cows of type-theory: ``modular typechecking''~\citep{Cardelli97}.
The idea behind modular typechecking is that each component of the program can be
typechecked once and for all, regardless of how and where it is used.
Since the methods of an object are generally interdependent, 
it seems clear that we cannot permit unconstrained changes to the type of a method
after the typecheck has been performed.
This runs contrary to the need to override methods to generalize functionality,  as we
have done in these examples.

Two paths are open to us.
The first,  exemplified by \citet*{ernst2001}, \citet*{bruce2003}, and \citet*{nystro2004},
is to devise clever restrictions that permit both useful overriding and early typechecking.  The second is to question the value of modular typechecking.

While modular typechecking seems obviously desirable for a subroutine library, 
or for a class that we are to \emph{instantiate}, its importance is less clear for a class that
we are going to \emph{inherit}.  
After all, inheritance lets us interfere with the internal connections between the 
methods of the inherited class, substituting arbitrary new code in place of
existing code.  
It is quite possible to break the inherited class by improper overriding: it is up to
the programmer who uses inheritance to understand the internal structure of the 
inherited class. 
In this context, running the typechecker over both the inherited code \emph{and} the
inheriting code seems like a small price to pay for the security that comes from being able 
to override method types and have the resulting composition be certified as type-safe.

\subsection{Further Extensions}

An issue we have not yet discussed is the difficulty of combining separate extensions, for example, 
including both output and exceptions in the expression language of Section~\ref{interp}.
Some form of multiple inheritance would seem to be indicated, but most attempts at multiple inheritance run into difficulties when the inherited parts conflict.
A particularly troublesome situation is the ``diamond problem'' \citep{Bracha1990,snyder1986a},
also known as ``fork-join inheritance'' \citep{sakkin1989};
this occurs when a class inherits a state component from the \emph{same} base class via multiple paths.

In our view, the most promising solution seems to be to restrict what can be multiply inherited.
For example, traits, as defined by \citet*{ducass2006}, consist of collections of methods that can be ``mixed into'' classes.  
Traits do not contain state; they instead access state declared elsewhere using
appropriate methods.  
The definition of a trait can also specify ``required methods'' that must be provided by the client.  
In exchange for these modest restrictions on expressiveness, traits avoid almost all of 
the problems of multiple inheritance.
Traits have been adopted by many recent languages; we are in the process of adding them to Grace.
Delegation may provide an alternative solution, but seems to be further from the mainstream of object-oriented language development.


One of the anonymous reviewers observed that \citeauthor{edward2005}' \textit{Subtext} \citeyearpar{edward2005} is
also based on the idea that humans are better at concrete thinking than at abstraction.
Subtext seeks to ``decriminalize copy and paste''~\citep{edward2005v} by supporting 
\textit{post hoc} abstraction.
Because Subtext programs are structures in a database rather than pieces of text,
the same program can be viewed in multiple ways.
Thus, whether someone reading a program sees a link to an abstraction or an embedded
implementation of that abstraction depends
on a check-mark on a style sheet, 
not on a once-and-for-all decision made by the writer of the program.
Moreover, the view can be changed in real time to suit the convenience of the reader.
Similar ideas were explored by \citet{BlacWODISEE}.

As we have shown in this article, inheritance can lead to highly-factored code.
This can be a double-edged sword. 
We have argued that the factoring is advantageous, 
because it lets the reader of a program digest it in 
small, easily understood chunks. 
But it can sometimes also be a disadvantage, 
because it leaves the reader with a less integrated view of the program. 
The ability to view a program in multiple ways can give us the best of both worlds, by 
allowing the chunks either to be viewed separately, or to be ``flattened'' into a single larger
chunk without abstraction boundaries. 
A full exploration of these ideas requires not only that we 
``step away from the ASR-33''~\citep{kamp2010}, but that we move away from
textual representations of programs altogether.

\subsection{Conclusion}
\label{conclusion}

Introducing a new concept by means of a series of examples is a technique as
old as pedagogy itself.
All of us were taught that way, and teach that way.  
We do not start instruction in mathematics by first explaining the abstractions of rings and fields, and then introducing arithmetic on the integers as a special case.
On the contrary: we introduce the abstractions only after the student is familiar with not one, but several, concrete examples.

As~\cite{banias2009} have written, 
\begin{quote}\small
the code that constitutes a program actually forms a higher-level, program-specific language. The symbols of the language are the abstractions of the program, and the grammar of the language is the set of (generally unwritten) rules about the allowable combinations of those abstractions.   As such, a program is both a language definition, and the only use of that language.  This specificity means that reading a never-before encountered program involves learning a new natural language
\end{quote}

It follows that when we write a program, we are teaching a new language.
Isn't it our duty to use all available technologies to improve our teaching?
So, next time you find yourself explaining a complex program by introducing a series of
simpler programs of increasing complexity, think how
that \emph{explanation}, as well as the final program, could be captured in your programming language. 
And if your chosen programming language does not provide the right tools, 
ask whether it could be improved.

\renewcommand\bibname{References}
\bibliographystyle{splncsnat}
\bibliography{references}

%
%
%
%
%
%
%

\end{document}